# LIMIT THEOREMS FOR QUANTUM WALKS ASSOCIATED WITH HADAMARD MATRICES

**CLEMENT AMPADU**

31 Carrolton Road
Boston, Massachusetts, 02132
U.S.A.
e-mail: drampadu@hotmail.com

**Abstract**

We study a one-parameter family of discrete-time quantum walk models on $Z$ and $Z^2$ associated with the Hadamard walk. Weak convergence in the long-time limit of all moments of the walker's pseudo-velocity on $Z$ and $Z^2$ is proved. *Symmetrization* on $Z$ and $Z^2$ is theoretically investigated, leading to the resolution of the Konno-Namiki-Soshi conjecture in the special case of symmetrization of the unbiased Hadamard walk on $Z$. A necessary condition for the existence of a phenomenon known as *localization* is given.

## 1. Introduction

The Hadamard walk plays a key role in studies of the quantum walk, thus the generalization of the Hadamard walk is one of the many fascinating challenges. As is well known, the simplest and well studied example of a quantum walk, [1] for example, is the Hadamard whose unitary matrix is defined by

$$H = \frac{1}{\sqrt{2}} \begin{bmatrix} 1 & 1 \\ 1 & -1 \end{bmatrix} \tag{1.1}$$

The dynamics of this walk corresponds to the symmetric random walk in the classical case. A generalization of (1.1) that leads to symmetric random walks and have been studied by a number of authors, including those in [2,3,4] is the following

$$H(\eta,\phi,\psi) = \frac{e^{i\eta}}{2} \begin{bmatrix} e^{i(\phi+\psi)} & e^{-i(\phi-\psi)} \\ e^{i(\phi-\psi)} & e^{-i(\phi+\psi)} \end{bmatrix} \qquad (1.2)$$

In this paper we study certain generalizations of the Hadamard walk in one- and two- dimensions and clarify theoretically the properties of *localization* and *symmetrization.* The paper begins in Section 2, where the stochastic process of the one- and two-dimensional quantum walk is introduced, there the generator of the process in the generalized Hadamard walk in both the one- and two- dimensional models are given, because the process of building the stochastic model is intimately related to the work of the authors in [5] we follow the convention of putting a superscript $T$ on the left throughout to denote the transpose of a vector/matrix. In Section 3 we prove convergence in the long-time limit of the walker's pseudo-velocity in one dimension, and do the same for the $x-$ and $y-$ components of the walker's pseudo-velocity in two dimension, since a common method of proof in the analysis of the long-time behavior of quantum walks whose time-evolution is given by unitary transformations is to diagonalize the time-evolution matrix, we use the method given by the authors in [5] which is essentially due to those in [2], and have been successfully used by a number of authors including those in [6, 7]. Due to the unitary nature of the time-evolution matrix, the convergence theorems are in a sense *weak* [8]. We also give in Section 3 analytic expressions for the probability distribution in both the one- and two- dimensional models, this first requires expressing the dynamics of the Hadamard walk considered in this paper in terms of difference equations [9,10, 11]. In Section 4 we study localization in the generalized Hadamard walk model. In either the one or two dimensional models we wish to answer the following question:

If say a quantum walker which could be a quantum particle exist at only at one site initially on the line in the one dimensional model or in the xy-plane in the two dimensional model, will the quantum walker remain trapped with high probability near the initial position?

The answer to this question has been investigated by various authors in varying contexts under the pseudonym "localization." For example Inui et.al [12] studied a generalized Hadamard walk in one dimension with three inner states and concluded that the quantum walker (quantum particle) is trapped near the origin with high probability. Watabe et.al [5] on the other hand were able to control localization around the origin for a one-parameter family of discrete-time quantum walk models on the square lattice, which included the Grover-walk which is related to the Grover's algorithm in computer science, as a special case. For the Grover-walk itself in two dimensions, Inui et.al [13] were able to show localization analytically, Chaobin Liu and Nelson Pentulante [14] were able to offer theoretical explanation for localization in the case of discrete quantum random walks on a linear lattice with two entangled coins. Following the localization studies by these and many other authors in the literature, in this section we offer theoretical criterion for localization with respect to the generalized Hadamard walk models considered in this paper. If the theoretical criterion holds then we should observe in computer simulations oscillatory behavior in the distribution of the pseudo-velocity in both the one and two dimensional models. In particular, as the time step increases a spike in the oscillatory behavior should be observed. However convergence of any moments given by either Theorem 3.5 (two dimensional model) or Theorem 3.2 (one-dimensional model) in this paper, imply if we smear out the oscillatory behavior, the averaged values of the distribution will be captured by the derived probability density function in both cases. Due to the expected dynamics as seen by previous authors we first give the criterion for localization anywhere on the line or in the plane by expressing the probability of localization in terms of the intensity of the Dirac Delta function, by making use of the stationary probability distribution in both dimensions. These criterions are given in Theorems 4.1 and 4.2 respectively. Due to the similar nature in proof of Theorem 4.2 to that of Theorem 4.1, we omit it, but remark that it involves invoking the Mean Value Property as it pertains to double integrals [15]. In Section 5, symmetry of the distribution in both the one and two dimensional models is considered theoretically, there the necessary and sufficient conditions for symmetrization is given. Along the way we settle the Konno-Namiki-Soshi conjecture as it relates to the symmetrization of the unbiased Hadamard walk in one dimension. Section 6 is devoted to the conclusions, there we leave the reader with an open problem.

## 2. The Quantum Walk Models

### 2.1 The One Dimensional Model

Let $R = \{x : x \in Z\}$, where $Z$ denotes the set of all integers $Z = \{\ldots, -2, -1, 0, 1, 2, \ldots\}$. Corresponding to the fact that there are two nearest-neighbor sites for each site $x \in R$, we assign a two-component wave function $\zeta(x,t) = \begin{pmatrix} \phi_1(x,t) \\ \phi_2(x,t) \end{pmatrix}$ to a quantum walker, each component of which is a complex function of location $x \in R$ and discrete time $t = 0,1,2,\ldots$. A quantum coin will be given by a $2 \times 2$ unitary matrix, $U = (U_{mn})_{m,n=1}^{2}$, and a spatial shift-operator on $R$ in the wave-number space $x' \in [-\pi, \pi)$ by the matrix $V(x') = \begin{pmatrix} e^{ix'} & 0 \\ 0 & e^{ix'} \end{pmatrix}$ where $i = \sqrt{-1}$. We assume that at the initial time $t = 0$ the walker is located at the origin with a two-component quidit $^T\Theta = (d_1 \;\; d_2) \in C^2$, where $C$ denotes the set of complex numbers, and $\sum_{i=1}^{2} |d_i|^2 = 1$.

**Definition 2.2 (Wave Function of the Walker):** Let $M(x') \equiv V(x')U$, the wave function of the walker at time $t$ is given by $\hat{\zeta}(x',t) = (M(x'))^t \Theta$, $t = 0,1,2,\ldots$ in the $x'$–space

Time evolution in $Z$ is obtained by performing the Fourier transformation $\zeta(x,t) = \int_{-\pi}^{\pi} \frac{dx'}{2\pi} e^{ix'x} \hat{\zeta}(x',t)$, where the inverse Fourier transform is given by $\hat{\zeta}(x',t) = \sum_{x \in Z} \zeta(x,t) e^{-ix'x}$.

**Definition 2.3 (Stochastic Process of One-Dimensional Quantum Walk):** Let $X_t$ be the x-coordinate of the position of the quantum walker at time t. The probability that we find the walker at $x \in R$ at time $t$ is given by $P(x,t) \equiv \text{Prob}(X_t = x) = {}^T\overline{\zeta}(x,t)\zeta(x,t)$, where the bar denotes complex conjugation.

The moment of $X_t$ is given by $\left\langle X_t^\alpha \right\rangle = \sum_{x \in Z} x^\alpha P(x,t) = \int_{-\pi}^{\pi} \frac{dx'}{2\pi} {}^T\overline{\zeta}(x',t) \left( i \frac{\partial}{\partial x'} \right)^\alpha \hat{\zeta}(x',t)$

for $\alpha = 0,1,2,\ldots$

## 2.2 The Two-Dimensional Model

Let $M = \{(x, y): x, y \in Z\}$, where $Z$ denotes the set of all integers $Z = \{\ldots, -2, -1, 0, 1, 2, \ldots\}$.

Corresponding to the fact that there are four nearest-neighbor sites for each site $(x, y) \in M$, we assign a four-component wave function $\Omega(x, y, t) = \begin{pmatrix} \varsigma_1(x,y,t) \\ \varsigma_2(x,y,t) \\ \varsigma_3(x,y,t) \\ \varsigma_4(x,y,t) \end{pmatrix}$ to a quantum walker, each component of which is a complex function of location $(x, y) \in M$ and discrete time $t = 0,1,2,\ldots$. A quantum coin will be given by a $4 \times 4$ unitary matrix, $B = (B_{st})_{s,t=1}^4$, and a spatial shift-operator on $M$ in the wave-number space $(m', n') \in [-\pi, \pi) \times [-\pi, \pi)$ by the matrix $T(m', n') = \begin{pmatrix} e^{im'} & 0 & 0 & 0 \\ 0 & e^{-im'} & 0 & 0 \\ 0 & 0 & e^{in'} & 0 \\ 0 & 0 & 0 & e^{-in'} \end{pmatrix}$, where

$i = \sqrt{-1}$. We assume that at the initial time $t = 0$ the walker is located at the origin with a four-

component quidit ${}^T \theta = (k_1 \ k_2 \ k_3 \ k_4) \in C^4$, where $C$ denotes the set of complex numbers, and

$\sum_{i=1}^{4} |k_i|^2 = 1$.

**Definition 2.4 (Wave Function of the Walker):** Let $Q(m',n') \equiv T(m',n')B$, the wave function of the walker at time $t$ is given by $\hat{\Omega}(m',n',t) = (Q(m',n'))^t \theta$, $t = 0,1,2,\ldots$ in the $(m',n')$–space

Time evolution in $M$ is obtained by performing the Fourier transformation

$$\Omega(x,y,t) = \int_{-\pi}^{\pi}\frac{dm'}{2\pi}\int_{-\pi}^{\pi}\frac{dn'}{2\pi}e^{i(m'x+n'y)}\widehat{\Omega}(m',n',t),$$ where the inverse Fourier transform is given by

$$\widehat{\Omega}(m',n',t) = \sum_{(x,y)\in M}\Omega(x,y,t)e^{-i(m'x+n'y)}.$$

**Definition 2.5 (Stochastic Process of Two-Dimensional Quantum Walk):** Let $X_t$ be the x-coordinate and let $Y_t$ be the y-coordinate of the position of the quantum walker at time t. The probability that we find the walker at $(x,y) \in M$ at time $t$ is given by

$$P(x,y,t) \equiv \Pr ob((X_t,Y_t) = (x,y)) = {}^T\overline{\Omega}(x,y,t)\Omega(x,y,t),$$ where the bar denotes complex conjugation. The joint moment of $X_t$ and $Y_t$ is given by

$$\left\langle X_t^\alpha Y_t^\beta \right\rangle \equiv \sum_{(x,y)\in M} x^\alpha y^\beta P(x,y,t) = \int_{-\pi}^{\pi}\frac{dm'}{2\pi}\int_{-\pi}^{\pi}\frac{dn'}{2\pi}\,{}^T\overline{\widehat{\Omega}}(m',n',t)\left(i\frac{\partial}{\partial m'}\right)^\alpha\left(i\frac{\partial}{\partial n'}\right)^\beta \widehat{\Omega}(m',n',t)$$ for

$\alpha, \beta = 0, 1, 2, \ldots$

## 2.3 Generalizations of the Hadamard Walk

In this paper we consider the following generalization of the Hadamard walk for the one-dimensional model

$$H^*(p,q) = \begin{bmatrix} \sqrt{p} & \sqrt{q} \\ \sqrt{q} & -\sqrt{p} \end{bmatrix}; \; q = 1-p \qquad (2.3.3)$$

where $p \in (0,1)$. In the two-dimensional model we will consider the following

$$H^{**}(p,q) = \begin{bmatrix} p & \sqrt{pq} & \sqrt{pq} & q \\ \sqrt{pq} & -p & q & -\sqrt{pq} \\ \sqrt{pq} & q & -p & -\sqrt{pq} \\ q & -\sqrt{pq} & -\sqrt{pq} & p \end{bmatrix}; \; q = 1-p \qquad (2.3.4)$$

where $p \in (0,1)$. When $p = \dfrac{1}{2}$, $H^*(p,q)$ reduces to the Hadamard matrix in equation (2.3.1), whilst $H^{**}(p,q)$ reduces to a two-dimensional generalization of (2.3.1). We close this section by noting that the generator of the process in the one-dimensional model is

$$M(x') = \begin{bmatrix} \sqrt{p}\, e^{ix'} & \sqrt{q}\, e^{ix'} \\ \sqrt{q}\, e^{-ix'} & -\sqrt{p}\, e^{-ix'} \end{bmatrix} \tag{2.3.5}$$

and in the two-dimensional model it is

$$Q(m',n') = \begin{bmatrix} p e^{im'} & \sqrt{pq}\, e^{im'} & \sqrt{pq}\, e^{im'} & q e^{im'} \\ \sqrt{pq}\, e^{-im'} & -p e^{-im'} & q e^{-im'} & -\sqrt{pq}\, e^{-im'} \\ \sqrt{pq}\, e^{in'} & q e^{in'} & -p e^{in'} & \sqrt{pq}\, e^{in'} \\ q e^{-in'} & -\sqrt{pq}\, e^{-in'} & -\sqrt{pq}\, e^{-in'} & p e^{-in'} \end{bmatrix} \tag{2.3.6}$$

In both (2.3.5) and (2.3.6), $q = 1 - p$ and $p \in (0,1)$.

### 3. Limit Distribution of Long-Time Behavior

#### 3.1 Weak Convergence in the One Dimensional Model

We first diagonalize the time-evolution matrix in the one dimensional model which is given by the matrix in equation (2.3.5). The eigenvalues of the matrix in equation (2.3.5) can be shown to be

$\mu_1 = e^{i\sigma(x')}$ and $\mu_2 = -e^{-i\sigma(x')}$ where $\sigma(x')$ is determined by the equation

$$\sin \sigma(x') = \sqrt{p}\, \sin(x') \tag{3.1.1}$$

The eigenvectors corresponding to the eigenvalues $\mu_j$, for $j = 1, 2$, are given by the following column vectors

$$\vec{h}_j(x') = N_j \begin{pmatrix} \dfrac{\sqrt{pq} + \sqrt{q}\mu_j e^{ix'}}{q} \\ 1 \end{pmatrix} \qquad (3.1.2)$$

where $N_j$ is an appropriate normalization factor. Define the $2\times 2$ unitary matrix by $P(x') = (\vec{h}_1 \ \vec{h}_2)$ where $\vec{h}_j(x')$ is the eigenvector corresponding to $\mu_j$. Put $D = \begin{pmatrix} \mu_1 & 0 \\ 0 & \mu_2 \end{pmatrix}$, then it follows that

$[P(x')]^{-1} M(x') P(x') = D$, and thus $M(x')$ is diagonalizable. Since $P(x')$ is unitary, it follows that $P(x')$ is invertible and that $[P(x')]^{-1} = \overline{{}^T P(x')}$. Now recall that the wave function of the walker at time $t$ in the one dimensional model is given by $\hat{\zeta}(x',t) = (M(x'))^t \Theta$, it follows from

$[P(x')]^{-1} M(x') P(x') = D$ that we can write $[M(x')]^t = [M(x') D (M(x'))^{-1}]^t$. By induction on $t$ we can show that $[M(x')]^t = [M(x') D (M(x'))^{-1}]^t = M(x') D^t [M(x')]^{-1}$. So the wave function of the walker at time $t$ in the one dimensional model can be written as $\hat{\zeta}(x',t) = M(x') D^t [M(x')]^{-1} \Theta$, using sigma notation we can show that the matrix $\hat{\zeta}(x',t) = M(x') D^t [M(x')]^{-1} \Theta$ can be written as

$\hat{\zeta}(x',t) = \sum_{j=1}^{2} \mu_j^t \vec{h}_j c_j(x')$ where $c_j(x') \equiv \overline{{}^T \vec{h}_j} \Theta$. It follows from $\hat{\zeta}(x',t) = \sum_{j=1}^{2} \mu_j^t \vec{h}_j c_j(x')$, that we have $\left(i\dfrac{\partial}{\partial x'}\right)^\alpha \hat{\zeta}(x',t) = \left(i\dfrac{\partial}{\partial x'}\right)^\alpha \left[e^{i\sigma(x')t} \vec{h}_1 c_1 + (-1)^t e^{-i\sigma(x')t} \vec{h}_2 c_2\right]$, for $\alpha = 1,2,\ldots$. By induction on

$\alpha$ we can show that $\left(i\dfrac{\partial}{\partial x'}\right)^\alpha \hat{\zeta}(x',t) = (-1)^{\alpha-1} \left[\dfrac{\partial}{\partial x'} \sigma(x')\right]^\alpha \left[(-1)^t \vec{h}_2 c_2 e^{i\sigma(x')t} - \vec{h}_1 c_1 e^{i\sigma(x')t}\right] t^\alpha$ for

$\alpha = 1,2,\ldots$. Since $P(x')$ is unitary and the eigenvectors $\vec{h}_j$ have been normalized it follows from the unitary of $P(x')$ that the vectors $\vec{h}_j$ are orthonormal since in particular $\overline{{}^T P(x')} P(x') = I_2$, where $I_2$ is the $2\times 2$ identity matrix, thus, $\overline{^T h_m(x')} h_{m'}(x') = \begin{cases} 1, & \text{if } m = m' \\ 0, & \text{if } m \neq m' \end{cases}$.

Now,

$$\overline{^T \hat{\zeta}(x',t)}\left(i\frac{\partial}{\partial x'}\right)^\alpha \hat{\zeta}(x',t) = \left[(-1)^t \overline{^T \vec{h}_2 \vec{c}_2} e^{-i\sigma(x')t} - \overline{^T \vec{h}_1 \vec{c}_1} e^{-i\sigma(x')t}\right](-1)^{\alpha-1}\left[\frac{\partial}{\partial x'}\sigma(x')\right]^\alpha \quad (3.1.3)$$
$$\left[(-1)^t \vec{h}_2 c_2 e^{i\sigma(x')t} - \vec{h}_1 c_1 e^{i\sigma(x')t}\right]t^\alpha$$

Since $\overline{^T h_m(x')} h_{m'}(x') = \begin{cases} 1, & \text{if } m = m' \\ 0, & \text{if } m \neq m' \end{cases}$, and noting that $|c_j|^2 = c_j \bar{c}_j$, it follows that we write the

expression in equation (3.1.3) as

$$\overline{^T \hat{\zeta}(x',t)}\left(i\frac{\partial}{\partial x'}\right)^\alpha \hat{\zeta}(x',t) = \left[\frac{\partial}{\partial x'}\sigma(x')\right]^\alpha \left[|c_2|^2 + |c_1|^2\right](-1)^{\alpha-1} t^\alpha \quad (3.1.4)$$

**Definition 3.1 (Pseudo-velocity of Quantum Walker):** For the one-dimensional model this is given as

$$V_t = \frac{X_t}{t}, \quad t = 1, 2, \ldots$$

Recall that $\langle X_t^\alpha \rangle = \int_{-\pi}^{\pi} \frac{dx'}{2\pi} \overline{^T \hat{\zeta}(x',t)}\left(i\frac{\partial}{\partial x'}\right)^\alpha \hat{\zeta}(x',t)$, however equation (3.1.4) above imply we have

$$\langle X_t^\alpha \rangle = \int_{-\pi}^{\pi} \frac{dx'}{2\pi} \{|c_2|^2 + |c_1|^2\}(-1)^{\alpha-1}\left[\frac{\partial}{\partial x'}\sigma(x')\right]^\alpha t^\alpha \quad (3.1.5)$$

From equation (3.1.1) we have $\sigma(x') = \arcsin\left(\sqrt{p}\sin(x')\right)$, and thus

$$\frac{\partial}{\partial x'}\sigma(x') = \frac{\sqrt{p}\cos(x')}{\sqrt{1 - p\sin^2 x'}} \quad (3.1.6)$$

From (3.1.5) and (3.1.6) we arrive at the main result, the convergence theorem in the long-time limit of the moment of the walker's pseudovelocity $\frac{X_t}{t}$.

**Theorem 3.2:** $\lim\limits_{t\to\infty}\left\langle\left(\dfrac{X_t}{t}\right)^\alpha\right\rangle = \int_{-\pi}^{\pi}\dfrac{dx'}{2\pi}\{|c_1|^2+|c_2|^2\}(-1)^{\alpha-1}\left[\dfrac{\sqrt{p}\cos(x')}{\sqrt{1-p\sin^2(x')}}\right]^\alpha$

### 3.2 Analytic Expression for the Probability Distribution in the One-Dimensional Model

The dynamics of the one-dimensional model considered herein can be expressed in terms of difference Equations giving the system of linear equations

$$\phi_1(x,t) = e^{ik}\left[\sqrt{p}\,\phi_1(x-1,t-1) + \sqrt{q}\,\phi_2(x-1,t-1)\right] \qquad (3.2.1)$$

$$\phi_2(x,t) = e^{ik}\left[\sqrt{q}\,\phi_1(x+1,t-1) - \sqrt{p}\,\phi_2(x+1,t-1)\right] \qquad (3.2.2)$$

for a given $\zeta(x,0) = \begin{pmatrix}d_1\\d_2\end{pmatrix}\delta_{x,0}$, where $k \in \Re$, $\delta_{x,0} = \begin{cases}1, \text{ if } x=0\\0, \text{ otherwise}\end{cases}$, $\phi_1$ and $\phi_2$ are the components of $\zeta(x,t)$, and $q = 1-p$. Now applying the Fourier transform $\hat{\zeta}(x',t) = \sum\limits_{x\in Z}\phi(x,t)e^{-ix'x}$ to (3.2.1) and (3.2.2) we obtain

$$\hat{\zeta}_1(x',t) = e^{ik}\left[\sqrt{p}\,e^{-ix'}\hat{\zeta}_1(x',t-1) + \sqrt{q}\,e^{-ix'}\hat{\zeta}_2(x',t-1)\right] \qquad (3.2.3)$$

$$\hat{\zeta}_2(x',t) = e^{ik}\left[\sqrt{q}\,e^{ix'}\hat{\zeta}_1(x',t-1) - \sqrt{p}\,e^{ix'}\hat{\zeta}_2(x',t-1)\right] \qquad (3.2.4)$$

In matrix form the system consisting of equations (3.2.3) and (3.2.4) can be written as

$\hat{\zeta}(x',t) = e^{ik}\,S(x')\,\hat{\zeta}(x',t-1)$, where $S(x') = \begin{pmatrix}\sqrt{p}\,e^{-ix'} & \sqrt{q}\,e^{-ix'}\\ \sqrt{q}\,e^{ix'} & -\sqrt{p}\,e^{ix'}\end{pmatrix}$. Now applying the Fourier transform $\hat{\zeta}(x',t) = \sum\limits_{x\in Z}\phi(x,t)e^{-ix'x}$ to the initial condition $\phi(x,0)$ we get the equivalent initial condition $\begin{pmatrix}\hat{\zeta}_1(x',0)\\\hat{\zeta}_2(x',0)\end{pmatrix} = \begin{pmatrix}\phi_1(0,0)\\\phi_2(0,0)\end{pmatrix}$. By induction on $t$ we can show that the wave function of the walker can be written as $\hat{\zeta}(x',t) = e^{itk} S^t(x')\hat{\zeta}(x',0)$. By way of the relation $e^{i\theta} = \cos\theta + i\sin\theta$ and properties of the Pauli matrices we can show that $S(x') = \begin{pmatrix} \sqrt{p}e^{-ix'} & \sqrt{q}e^{-ix'} \\ \sqrt{q}e^{ix'} & -\sqrt{p}e^{ix'} \end{pmatrix}$ admits the following decomposition $S(x') = \cos(\theta(x'))I + i\sin(\theta(x'))\bar{c}(x')\cdot\bar{\sigma}$, where $\theta$ and $\bar{c}$ are real functions of $x'$, and the matrix vector $\bar{\sigma}$ has Pauli components $\bar{\sigma}_1 = \begin{bmatrix} 0 & 1 \\ 1 & 0 \end{bmatrix}$, $\bar{\sigma}_2 = \begin{bmatrix} 0 & -i \\ i & 0 \end{bmatrix}$, $\bar{\sigma}_3 = \begin{bmatrix} 1 & 0 \\ 0 & -1 \end{bmatrix}$, where $I$ is the $2\times 2$ identity matrix. The decomposition $S(x') = \cos(\theta(x'))I + i\sin(\theta(x'))\bar{c}(x')\cdot\bar{\sigma}$ is exponential and implies we can write $S^t(x') = \cos(t\theta(x'))I + i\sin(t\theta(x'))\bar{c}(x')\cdot\bar{\sigma}$.

**Definition 3.3:** The Chebyshev polynomial $T_n(x)$ of the first kind is a polynomial in $x$ of degree $n$, defined by the relation $T_n(x) = \cos(n\theta)$ when $x = \cos(\theta)$, and the Chebyshev polynomial $U_n(x)$ of the second kind is a polynomial in $x$ of degree $n$ defined by the relation $U_n(x) = \dfrac{\sin(n+1)\theta}{\sin\theta}$ when $x = \cos(\theta)$ [16].

Using Definition 3.3 we can can write $S^t(x') = T_t(\cos(\theta(x')))I + U_{t-1}(\cos(\theta(x')))i\sin(\theta(x'))\bar{c}(x')\cdot\bar{\sigma}$.

In Mason et.al [16] it is shown that $U_n(x) = 2xU_{n-1}(x) - U_{n-2}(x)$ and $2T_n(x) = U_n(x) - U_{n-2}(x)$. From both of these relations we can deduce $T_n(x) = U_n(x) - xU_{n-1}(x)$, from which it follows that we can write

$$S^t(x') = U_t(\cos(\theta(x')))I - U_{t-1}(\cos(\theta(x')))[\cos(-\theta(x'))I + i\sin(-\theta(x'))\bar{c}(x').\bar{\sigma}] \qquad (3.2.5)$$

However, we notice that the expression within the square brackets in equation (3.2.5) can be written using exponential notation as $\left[e^{i\theta(x')\bar{c}(x').\bar{\sigma}}\right]^{-1} = [S(x')]^{-1}$. So we can write equation (3.2.5) as

$$S^t(x') = U_t(\cos(\theta(x')))I - U_{t-1}(\cos(\theta(x')))[S(x')]^{-1} \qquad (3.2.6)$$

Following Fuss et.al [9] we can also determine the function $\cos(\theta(x'))$ in terms of the components of $S(x')$ by defining the inner product $(A, B) = \frac{1}{2}Tr(AB)$, where $Tr(.)$ denotes the trace function, on the vector space of two by two unitary matrices and hence obtain an inner product space with $\{I, \sigma_1, \sigma_2, \sigma_3\}$ as an orthonormal basis. The co-efficient of the identity matrix on the right side of $S(x') = \cos(\theta(x'))I + i\sin(\theta(x'))\vec{c}(x') \cdot \vec{\sigma}$ is $\cos(\theta(x'))$. To determine the coefficient of the identity matrix on the left hand side of $S(x') = \cos(\theta(x'))I + i\sin(\theta(x'))\vec{c}(x') \cdot \vec{\sigma}$, we take the inner product,

$(I, S(x')) = \frac{1}{2}Tr(I\,S(x'))$, which gives $-i\sqrt{p}\sin(x')$. We saw earlier that the wave function of the walker at an arbitrary time $t$ is given by $\hat{\zeta}(x',t) = e^{itk}S^t(x')\hat{\zeta}(x',0)$, thus between two times $t_0$ and $t_1$, $\hat{\zeta}(x',t) = e^{itk}S^t(x')\hat{\zeta}(x',0)$ has the equivalent interpretation $T(t_1, t_0) = e^{i(t_1-t_0)k}S^{(t_1-t_0)}(x')$. So we can write

$$T(t,0) = e^{itk}S^t(x')$$
$$= e^{itk}\left[U_t(\cos(\theta(x')))I - U_{t-1}(\cos(\theta(x')))[S(x')]^{-1}\right] \quad (3.2.7)$$
$$= e^{itk}\left[U_t(-i\sqrt{p}\sin(x'))I - U_{t-1}(-i\sqrt{p}\sin(x'))[S(x')]^{-1}\right]$$

From $\hat{\zeta}(x',t) = e^{itk}S^t(x')\hat{\zeta}(x',0)$, $T(t_1, t_0) = e^{i(t_1-t_0)k}S^{(t_1-t_0)}(x')$, and equation (3.2.7) we can write

$$\hat{\zeta}(x',t) = e^{itk}\left[U_t(-i\sqrt{p}\sin(x'))I - U_{t-1}(-i\sqrt{p}\sin(x'))[S(x')]^{-1}\right]\hat{\zeta}(x',0) \quad (3.2.8)$$

In order to write equation (3.2.8) in terms of $\hat{\zeta}_1(x',t)$ and $\hat{\zeta}_2(x',t)$ we first note that

$$[S(x')]^{-1} = \begin{bmatrix} \sqrt{p}e^{ix'} & \sqrt{q}e^{-ix'} \\ \sqrt{q}e^{ix'} & -\sqrt{p}e^{ix'} \end{bmatrix}, \hat{\zeta}(x',t) = \begin{pmatrix} \hat{\zeta}_1(x',t) \\ \hat{\zeta}_2(x',t) \end{pmatrix}, \zeta(x',0) = \begin{pmatrix} \hat{\zeta}_1(x,0) \\ \hat{\zeta}_2(x,0) \end{pmatrix} = \begin{pmatrix} \phi_1(0,0) \\ \phi_2(0,0) \end{pmatrix} = \begin{pmatrix} d_1 \\ d_2 \end{pmatrix},$$

thus we have,

$$\hat{\zeta}_1(x',t) = e^{itk}\left[U_t(-i\sqrt{p}\sin(x'))d_1 - \left(\sqrt{p}e^{ix'}d_1 + \sqrt{q}e^{-ix'}d_2\right)U_{t-1}(-i\sqrt{p}\sin(x'))\right] \quad (3.2.9)$$

$$\hat{\zeta}_2(x',t) = e^{itk}\left[U_t(-i\sqrt{p}\sin(x'))d_2 + \left(\sqrt{p}e^{ix'}d_2 - \sqrt{q}e^{-ix'}d_1\right)U_{t-1}(-i\sqrt{p}\sin(x'))\right] \qquad (3.2.10)$$

Using the definition $u_t(-i\sqrt{p}:x) = \dfrac{1}{2\pi}\displaystyle\int_{-\pi}^{\pi} U_t(-i\sqrt{p}\sin(x'))e^{ixx'}dx'$, and applying the inverse Fourier

transform to equation (3.2.9) and equation (3.2.10) respectively we have

$$\phi_1(x,t) = e^{itk}\left[d_1 u_t\left(-i\sqrt{p}:x\right) - \sqrt{p}d_1 u_{t-1}(-i\sqrt{p}:x+1) - \sqrt{q}d_2 u_{t-1}(-i\sqrt{p}:x-1)\right] \qquad (3.2.11)$$

$$\phi_2(x,t) = e^{itk}\left[d_2 u_t\left(-i\sqrt{p}:x\right) + \sqrt{p}d_2 u_{t-1}(-i\sqrt{p}:x+1) - \sqrt{q}d_1 u_{t-1}(-i\sqrt{p}:x-1)\right] \qquad (3.2.12)$$

Since $\zeta(x,t) = \begin{pmatrix} \phi_1(x,t) \\ \phi_2(x,t) \end{pmatrix} \in C^2$ gives the time evolution of the quantum random walk for discrete times

$t \geq 0$ on a line $x \in Z$. We note that for a quantum particle starting from the initial state $\begin{pmatrix} d_1 \\ d_2 \end{pmatrix}$,

equations (3.2.11) and (3.2.12) imply that the quantum walk probability density is given by

$|\phi_1(x,t)|^2 + |\phi_2(x,t)|^2$. Write $P(x,t) = |\phi_1(x,t)|^2 + |\phi_2(x,t)|^2$, then from equations (3.2.11) and (3.2.12)

we have that

$$P(x,t) = \left|d_1 u_t\left(-i\sqrt{p}:x\right) - \sqrt{p}d_1 u_{t-1}(-i\sqrt{p}:x+1) - \sqrt{q}d_2 u_{t-1}(-i\sqrt{p}:x-1)\right|^2$$
$$+$$
$$\left|d_2 u_t\left(-i\sqrt{p}:x\right) + \sqrt{p}d_2 u_{t-1}(-i\sqrt{p}:x+1) - \sqrt{q}d_1 u_{t-1}(-i\sqrt{p}:x-1)\right|^2 \qquad (3.2.13)$$

To write equation (3.2.13) in closed form we use the power series method of [9]. According to

the authors we can write

$$U_t(y) = \sum_{m=0}^{\left[\frac{t}{2}\right]} (-1)^m \binom{t-m}{m}(2y)^{t-2m} \qquad (3.2.14)$$

where $\binom{t-m}{m} = \dfrac{(t-m)!}{m!(t-2m)!}$ for $0 \leq m \leq t-m$, and $\left[\dfrac{t}{2}\right]$ is the floor of $\dfrac{t}{2}$.

If we put $2y = \sqrt{p}(e^{-ix'} - e^{ix'})$ and expand the powers using the binomial theorem, it follows we can write

$$U_t(-i\sqrt{p}\sin(x')) = \sum_{m=0}^{\left[\frac{t}{2}\right]}(-1)^m \binom{t-m}{m}(\sqrt{p})^{t-2m}\sum_{k=0}^{t-2m}\binom{t-2m}{k}(-1)^k e^{-ix'(t-2(m+k))} \quad (3.2.15)$$

To put equation (3.2.15) in a suitable form, we change the index of summation in the second sum by putting $j = m+k$, it follows we can then write (3.2.15) as

$$U_t(-i\sqrt{p}\sin(x')) = \sum_{m=0}^{\left[\frac{t}{2}\right]}(-1)^m \binom{t-m}{m}(\sqrt{p})^{t-2m}\sum_{j=0}^{t}\binom{t-2m}{j-m}(-1)^k e^{-ix'(t-2j)} \quad (3.2.16)$$

where the property $\binom{q}{r} = 0$ is to be used when $r < 0$ or $r > q$. Reordering the summation in equation (3.2.16) gives

$$U_t(-i\sqrt{p}\sin(x')) = \sum_{j=0}^{t}\left[\sum_{m=0}^{\left[\frac{t}{2}\right]}(-1)^{m+k}\binom{t-m}{m}\binom{t-2m}{j-m}(\sqrt{p})^{t-2m}\right]e^{-ix'(t-2j)} \quad (3.2.17)$$

Let $P_{t-2j}^t(-i\sqrt{p})$ denote the expression in square brackets in equation (3.2.17), then

$$U_t(-i\sqrt{p}\sin(x')) = \sum_{j=0}^{t} P_{t-2j}^t(-i\sqrt{p})e^{-ix'(t-2j)} \quad (3.2.18)$$

Recall that $U_t(-i\sqrt{p}\sin(x'))$ has inverse Fourier transform

$u_t(-i\sqrt{p}:x) = \frac{1}{2\pi}\int_{-\pi}^{\pi}U_t(-i\sqrt{p}\sin(x'))e^{ixx'}dx'$. However by (3.2.18) we can write

$$u_t(-i\sqrt{p}:x) = \frac{1}{2\pi}\int_{-\pi}^{\pi}\sum_{j=0}^{t}P_{t-2j}^t(-i\sqrt{p})e^{ix'(x-(t-2j))}dx' \quad (3.2.19)$$

We see from equation (3.2.19) that $u_t(-i\sqrt{p}:x) = \sum_{j=0}^{t}P_{t-2j}^t(-i\sqrt{p})$ provided that $x = t-2j$, and $u_t(-i\sqrt{p}:x) = 0$ provided that $x \neq t-2j$. Thus in terms of the Kronecker delta we can write

$$u_t\left(-i\sqrt{p}:x\right)=\sum_{j=0}^{t}P_{t-2j}^{t}\left(-i\sqrt{p}\right)\delta_{x,t-2j},$$ where $P_{t-2j}^{t}\left(-i\sqrt{p}\right)$ is the expression inside the square brackets in equation (3.2.17). The main result of this section is now ready and we summarize as follows:

**Theorem 3.4:** The quantum probability density function for the generalized hadamard walk in one-dimension is given by

$$P(x,t)=\left|d_1u_t\left(-i\sqrt{p}:x\right)-\sqrt{p}d_1u_{t-1}(-i\sqrt{p}:x+1)-\sqrt{q}d_2u_{t-1}(-i\sqrt{p}:x-1)\right|^2$$
$$+$$
$$\left|d_2u_t\left(-i\sqrt{p}:x\right)+\sqrt{p}d_2u_{t-1}(-i\sqrt{p}:x+1)-\sqrt{q}d_1u_{t-1}(-i\sqrt{p}:x-1)\right|^2$$

where $u_t\left(-i\sqrt{p}:x\right)=\sum_{j=0}^{t}P_{t-2j}^{t}\left(-i\sqrt{p}\right)\delta_{x,t-2j}$, and $P_{t-2j}^{t}\left(-i\sqrt{p}\right)$ is the expression in square brackets in equation (3.2.17)

### 3.3 Weak Convergence in the Two Dimensional Model

We first note that $H^{**}(p,q)=H^{*}(p,q)\otimes H^{*}(p,q)$, where $H^{*}(p,q)$ is given by equation (2.3.3). From the matrix in equation (2.3.5) one can deduce that the eigenvalues and the eigenvectors of $H^{**}(p,q)$ are the *product* of the eigenvalues and the eigenvectors of the matrices $M\left(\frac{m'+n'}{2}\right)$ and $M\left(\frac{m'-n'}{2}\right)$, where the matrices $M\left(\frac{m'+n'}{2}\right)$ and $M\left(\frac{m'-n'}{2}\right)$ come from the matrix in equation (2.3.5) upon making the appropriate substitution, then using the method of Section 3.1 the convergence theorem in the long –time limit of the joint moment of the walker's pseudo-velocity $\frac{X_t}{t}$ and $\frac{Y_t}{t}$ Is given by the following.

**Theorem 3.5:**

$$\lim_{t \to \infty} \left\langle \left(\frac{X_t}{t}\right)^\alpha \left(\frac{Y_t}{t}\right)^\beta \right\rangle =$$

$$\iint_{[-\pi,\pi)^2} \left( \frac{-\sqrt{p}\cos\left(\frac{m'+n'}{2}\right)}{2\sqrt{1-p\sin^2\left(\frac{m'+n'}{2}\right)}} - \frac{\sqrt{p}\cos\left(\frac{m'-n'}{2}\right)}{2\sqrt{1-p\sin^2\left(\frac{m'-n'}{2}\right)}} \right)^\alpha \left( \frac{-\sqrt{p}\cos\left(\frac{m'+n'}{2}\right)}{2\sqrt{1-p\sin^2\left(\frac{m'+n'}{2}\right)}} + \frac{\sqrt{p}\cos\left(\frac{m'-n'}{2}\right)}{2\sqrt{1-p\sin^2\left(\frac{m'-n'}{2}\right)}} \right)^\beta$$
$$|c_1(m',n')|^2 \frac{dm'}{2\pi} \frac{dn'}{2\pi}$$

$$+ \iint_{[-\pi,\pi)^2} \left( \frac{-\sqrt{p}\cos\left(\frac{m'+n'}{2}\right)}{2\sqrt{1-p\sin^2\left(\frac{m'+n'}{2}\right)}} + \frac{\sqrt{p}\cos\left(\frac{m'-n'}{2}\right)}{2\sqrt{1-p\sin^2\left(\frac{m'-n'}{2}\right)}} \right)^\alpha \left( \frac{-\sqrt{p}\cos\left(\frac{m'+n'}{2}\right)}{2\sqrt{1-p\sin^2\left(\frac{m'+n'}{2}\right)}} - \frac{\sqrt{p}\cos\left(\frac{m'-n'}{2}\right)}{2\sqrt{1-p\sin^2\left(\frac{m'-n'}{2}\right)}} \right)^\beta$$
$$|c_2(m',n')|^2 \frac{dm'}{2\pi} \frac{dn'}{2\pi}$$

$$+ \iint_{[-\pi,\pi)^2} \left( \frac{-\sqrt{p}\cos\left(\frac{m'-n'}{2}\right)}{2\sqrt{1-p\sin^2\left(\frac{m'-n'}{2}\right)}} + \frac{\sqrt{p}\cos\left(\frac{m'+n'}{2}\right)}{2\sqrt{1-p\sin^2\left(\frac{m'+n'}{2}\right)}} \right)^\alpha \left( \frac{\sqrt{p}\cos\left(\frac{m'-n'}{2}\right)}{2\sqrt{1-p\sin^2\left(\frac{m'-n'}{2}\right)}} + \frac{\sqrt{p}\cos\left(\frac{m'+n'}{2}\right)}{2\sqrt{1-p\sin^2\left(\frac{m'+n'}{2}\right)}} \right)^\beta$$
$$|c_3(m',n')|^2 \frac{dm'}{2\pi} \frac{dn'}{2\pi}$$

$$+ \iint_{[-\pi,\pi)^2} \left( \frac{\sqrt{p}\cos\left(\frac{m'+n'}{2}\right)}{2\sqrt{1-p\sin^2\left(\frac{m'+n'}{2}\right)}} + \frac{\sqrt{p}\cos\left(\frac{m'-n'}{2}\right)}{2\sqrt{1-p\sin^2\left(\frac{m'-n'}{2}\right)}} \right)^\alpha \left( \frac{\sqrt{p}\cos\left(\frac{m'+n'}{2}\right)}{2\sqrt{1-p\sin^2\left(\frac{m'+n'}{2}\right)}} - \frac{\sqrt{p}\cos\left(\frac{m'-n'}{2}\right)}{2\sqrt{1-p\sin^2\left(\frac{m'-n'}{2}\right)}} \right)^\beta$$
$$|c_4(m',n')|^2 \frac{dm'}{2\pi} \frac{dn'}{2\pi}$$

### 3.4 Analytic Expression for the Probability Distribution in the Two-Dimensional Model

Note that the dynamics of the two-dimensional Hadamard model considered in this paper can be expressed in terms of difference equations giving the system of linear equations

$$\phi_1(x,y,t) = e^{ik}\left\{p\phi_1(x-1,y,t-1) + \sqrt{pq}\phi_2(x-1,y,t-1) + \sqrt{pq}\phi_3(x-1,y,t-1) + q\phi_4(x-1,y,t-1)\right\}$$
(3.4.1)

$$\phi_2(x,y,t) = e^{ik}\left\{\sqrt{pq}\,\phi_1(x+1,y,t-1) - p\,\phi_2(x+1,y,t-1) + q\,\phi_3(x+1,y,t-1) - \sqrt{pq}\,\phi_4(x+1,y,t-1)\right\}$$
(3.4.2)

$$\phi_3(x,y,t) = e^{ik}\left\{\sqrt{pq}\,\phi_1(x,y-1,t-1) + q\,\phi_2(x,y-1,t-1) - p\,\phi_3(x,y-1,t-1) - \sqrt{pq}\,\phi_4(x,y-1,t-1)\right\}$$
(3.4.3)

$$\phi_4(x,y,t) = e^{ik}\left\{q\,\phi_1(x,y+1,t-1) - \sqrt{pq}\,\phi_2(x,y+1,t-1) - \sqrt{pq}\,\phi_3(x,y+1,t-1) + p\,\phi_4(x,y+1,t-1)\right\}$$
(3.4.4)

for a given $\Omega(x,y,0) = \begin{pmatrix} q_1 \\ q_2 \\ q_3 \\ q_4 \end{pmatrix} \in C^4$ understood when $x = y = 0$, where $C$ is the set of complex numbers and $\sum_{j=1}^{4}|q_j|^2 = 1$, $k \in R$, $q = 1-p$ and $\phi_j$ for $j = 1,2,3,4$ are the components of

$\Omega(x,y,t)$, then using the method of Section 3.2 we get the probability distribution for the generalized Hadamard walks as the following.

**Theorem 3.6:** The quantum walk probability density function is given by $P(x,y,t) = \sum_{i=1}^{4}|\phi_i(x,y,t)|^2$, where $\phi_1, \phi_2, \phi_3, \phi_4$ are given by

$$\phi_1(x,y,t) = e^{itk}\left\{q_1 u_t(ip:(x,y)) - q_1 p u_{t-1}(ip:(x+1,y)) - q_2\sqrt{pq}\,u_{t-1}(ip:(x,y+1)) - q_3\sqrt{pq}\,u_{t-1}(ip:(x,y-1)) - q_4\,q\,u_{t-1}(ip:(x-1,y))\right\}$$

$$\phi_2(x,y,t) = e^{itk}\left\{q_2 u_t(ip:(x,y)) - q_1\sqrt{pq}\,u_{t-1}(ip:(x+1,y)) + q_2 p u_{t-1}(ip:(x+1,y)) - q_3 q u_{t-1}(ip:(x,y-1)) + q_4\sqrt{pq}\,u_{t-1}(ip:(x,y-1))\right\}$$

$$\phi_3(x,y,t) = e^{itk}\left\{q_3 u_t(ip:(x,y)) - q_1\sqrt{pq}\,u_{t-1}(ip:(x+1,y)) - q_2 q u_{t-1}(ip:(x,y+1)) + q_3 p u_{t-1}(ip:(x+1,y)) + q_4\sqrt{pq}\,u_{t-1}(ip:(x,y+1))\right\}$$

$$\phi_4(x,y,t) = e^{itk}\left\{q_4 u_t(ip:(x,y)) - q_1 q u_{t-1}(ip:(x+1),y)) + q_2\sqrt{pq}\,u_{t-1}(ip:(x+1,y)) + q_3\sqrt{pq}\,u_{t-1}(ip:(x+1,y)) - q_4 p u_{t-1}(ip:(x+1,y))\right\}$$

and $u_t(ip:(x,y))$ in the expressions for $\phi_1, \phi_2, \phi_3, \phi_4$ is given by

$$u_t(ip:(x,y)) = \begin{cases} \sum_{j=0}^{t-k} Q(p), & \text{if } (x,y) = (-q, t-k-2j) \\ 0, & \text{otherwsie} \end{cases}, \text{ where}$$

$$Q(p) = \sum_{m=0}^{\left[\frac{t}{2}\right]} \sum_{k=0}^{t-2m} \sum_{q=0}^{k} \begin{bmatrix} t \\ m \end{bmatrix} p^{t-2m} \binom{t-k-2m}{j-m} \binom{t-2m}{k} \binom{k}{q} (-1)^{t-2m-v-q} \text{ , and}$$

$$\begin{bmatrix} t \\ m \end{bmatrix} = (-1)^m \binom{t-m}{m} \text{ and } \binom{t-m}{m} \text{ is the binomial coefficient as defined in equation (3.2.14).}$$

### 4. Localization in the Generalized Hadamard Model: Theoretical Investigation

To give the criterion we first give the probability of localization on the line in the one dimensional model.

**Theorem 4.1:** Let $P(x) = \lim_{t \to \infty} P(x,t)$, then the probability of localization at $x = x_0$ is given by

$$P(x_0) = \int_{-\infty}^{\infty} P(x) \delta(x - x_0) dx$$, where $P(x,t)$ is the quantum probability density function for the generalized hadamard walk in one dimension as given by Theorem 3.4, and $\delta(x)$ is the one-dimensional Dirac Delta function.

**Proof:** We make use of the following characterization of the Dirac Delta function which is available in [17]. Let $\varepsilon > 0$ be given. The Dirac Delta function, $\delta(x)$, can be defined as the limit of a sequence of discontinuous functions $\delta_\varepsilon(x)$, where

$$\delta_\varepsilon(x) = \begin{cases} \frac{1}{2\varepsilon}; |x| < \varepsilon \\ 0; |x| > \varepsilon \end{cases}$$

From this definition we see that by shifting $\delta_\varepsilon(x)$ to the right $x_0$ units we can define $\delta(x - x_0)$ as the limit of a sequence of discontinuous functions $\delta_\varepsilon(x - x_0)$ defined as

$$\delta_\varepsilon(x - x_0) = \begin{cases} \frac{1}{2\varepsilon}; |x - x_0| < \varepsilon \\ 0; |x - x_0| > \varepsilon \end{cases}$$. Each $\delta_\varepsilon(x - x_0)$ has unit area under the curve and

$\lim_{\varepsilon \to 0} \delta_\varepsilon (x - x_0) = 0$, if $x \neq x_0$. However, on formally interchanging the limit process with

integration, we obtain $\int_{-\infty}^{\infty} \delta_\varepsilon (x - x_0) dx = \lim_{\varepsilon \to 0} \int_{-\infty}^{\infty} \delta_\varepsilon (x - x_0) dx = 1$. Since $P(x) = \lim_{t \to \infty} P(x,t)$ is a

continuous function by virtue of the continuity of the quantum probability density function

$P(x,t)$, it follows upon using the mean value theorem for integrals before passing to the limit

that $P(x_0) = \int_{-\infty}^{\infty} P(x) \delta(x - x_0) dx$ and the proof is finished.

In the two dimensional model the probability of localization is given by the following.

**Theorem 4.2:** Let $P(x, y) = \lim_{t \to \infty} P(x, y, t)$, then the probability of localization at $(x, y) = (x_0, y_0)$ is given by $P(x_0, y_0) = \int_{-\infty}^{\infty} \int_{-\infty}^{\infty} P(x, y) \delta(x - x_0, y - y_0) dx dy$, where $P(x, y, t)$ is the quantum walk probability density function for the generalized Hadamard walk in two dimensions as given by Theorem 3.6, and $\delta(x, y)$ is the two-dimensional Dirac Delta function.

**Definition 4.3:** We say localization has occurred in either the one or two dimensional models if $P(x_0)$ or

$P(x_0, y_0)$ is *sufficiently large* within the confines of the interval $[0,1]$.

## 5. Symmetry of the Distribution: Theoretical Investigation

In this section we give rigorous results on the symmetry of the distribution for both the one and two dimensional models considered in this paper. We follow closely the ideas in the paper of Konno et.al [18].

### 5.1 Necessary and Sufficient Condition in the One-Dimensional Model

Recall that in the one-dimensional model that the Hadamard transformation is given by

$H^*(p,q) = \begin{bmatrix} \sqrt{p} & \sqrt{q} \\ \sqrt{q} & -\sqrt{p} \end{bmatrix}$. $H^*(p,q)$ acts on chirality states, which signifies the direction of motion of the particle. In the one-dimensional model the particle can either move left or right in the direction of the $x-axis$ depending on its chirality state. In particular, $H^*(p,q)$ acts on the chirality states $|L\rangle$ and $|R\rangle$, where $L$ and $R$ refer to the right and left chirality states respectively, and $|L\rangle = \begin{bmatrix} 1 \\ 0 \end{bmatrix}$ and $|R\rangle = \begin{bmatrix} 0 \\ 1 \end{bmatrix}$. In particular $H^*(p,q)$ acts on the two chirality states via: $|L\rangle \mapsto \sqrt{p}|L\rangle + \sqrt{q}|R\rangle$, $|R\rangle \mapsto \sqrt{q}|L\rangle - \sqrt{p}|R\rangle$, and thus

$H^*(p,q)|L\rangle = \sqrt{p}|L\rangle + \sqrt{q}|R\rangle$ and $H^*(p,q)|R\rangle = \sqrt{q}|L\rangle - \sqrt{p}|R\rangle$. Now define the

matrices $P$ and $Q$ as follows $P = \sqrt{p}\begin{bmatrix} 1 & 0 \\ 0 & 0 \end{bmatrix} + \sqrt{q}\begin{bmatrix} 0 & 1 \\ 0 & 0 \end{bmatrix}$ and

$Q = \sqrt{q}\begin{bmatrix} 0 & 0 \\ 1 & 0 \end{bmatrix} + \sqrt{p}\begin{bmatrix} 0 & 0 \\ 0 & -1 \end{bmatrix}$ with $H^*(p,q) = P + Q$. Note that $P$ represents left moving particle and $Q$ represents right moving particle. Now we define the dynamics of the Hadamard walk in one dimension using $P$ and $Q$. To do so we define the following $(2N+1) \times (2N+1)$ matrix $\overline{H}_N^* : (C^2)^{2N+1} \to (C^2)^{2N+1}$ where $C$ is the set of complex numbers, $P$ and $Q$ are as above, and $0_{2\times 2} = \begin{bmatrix} 0 & 0 \\ 0 & 0 \end{bmatrix}$

$$\overline{H}_N^* = \begin{bmatrix} 0_{2\times 2} & P & 0_{2\times 2} & \cdots & & \cdots & 0_{2\times 2} & Q \\ Q & 0_{2\times 2} & P & 0_{2\times 2} & \cdots & & \cdots & 0_{2\times 2} \\ 0_{2\times 2} & Q & 0_{2\times 2} & P & 0_{2\times 2} & \cdots & & 0_{2\times 2} \\ \vdots & \ddots & \ddots & \ddots & \ddots & \ddots & \ddots & \vdots \\ 0_{2\times 2} & \cdots & 0_{2\times 2} & Q & 0_{2\times 2} & P & & 0_{2\times 2} \\ 0_{2\times 2} & \cdots & & \cdots & 0_{2\times 2} & Q & 0_{2\times 2} & P \\ P & 0_{2\times 2} & \cdots & & \cdots & 0_{2\times 2} & Q & 0_{2\times 2} \end{bmatrix}$$

Now let $\zeta_x^t(\Theta) = \begin{pmatrix} \phi_{1,x}^t(\Theta) \\ \phi_{2,x}^t(\Theta) \end{pmatrix} = \phi_{1,x}^t(\Theta)|L\rangle + \phi_{2,x}^t(\Theta)|R\rangle \in C^2$ be the two component vector of amplitudes of the particle being at site $x$ and at time $t$ with the chirality being left (upper component) and right (lower component), and

$\zeta^t(\Theta) =^T [\zeta_{-N}^t(\Theta), \zeta_{-(N-1)}^t(\Theta), \ldots, \zeta_N^t(\Theta)] \in (C^2)^{2N+1}$ be the quibit states at time $t$, where $T$ represents the transposed operator. Here the initial quibit state is given by

$$\zeta^0(\Theta) = [\overbrace{0_{2\times 1}, \ldots 0_{2\times 1}}^{N}, \Theta, \overbrace{0_{2\times 1}, \ldots 0_{2\times 1}}^{N}] \in (C^2)^{2N+1}, \text{ where } 0_{2\times 1} = \begin{bmatrix} 0 \\ 0 \end{bmatrix}, \Theta = \begin{bmatrix} d_1 \\ d_2 \end{bmatrix}, \text{ with}$$

$\sum_{i=1}^{2}|d_i|^2 = 1$. The time evolution of the one dimensional Hadamard walk considered in this paper is then given by $(\overline{H}_N^* \zeta^t(\Theta))_x = Q\zeta_{x-1}^t(\Theta) + P\zeta_{x+1}^t(\Theta)$, where $(\zeta^t(\Theta))_x = \zeta_x^t(\Theta)$.

Since $P$ and $Q$ satisfy $PP^* = P^*P + Q^*Q = I$ and $PQ^* = QP^* = Q^*P = P^*Q = 0_{2\times 2}$, where the * represents the adjoint operator, it follows that the matrix $\overline{H}_N^*$ is also a unitary matrix. Now define the set of initial quibit states by $\Phi = \left\{ \Theta = \begin{bmatrix} d_1 \\ d_2 \end{bmatrix} \in C^2 : \sum_{i=1}^{2}|d_i|^2 = 1 \right\}$ and define the probability distribution of generalized Hadamard walk in one dimension by

$P(X_t^\Theta = x) = |\zeta_x^t(\Theta)|^2$.

To study the symmetry of the distribution of $H^*\left(\frac{1}{2}, \frac{1}{2}\right)$ Konno et.al [18] introduced the following classes of initial quibit states:

$\Phi_\perp = \left\{ \Theta = \begin{bmatrix} d_1 \\ d_2 \end{bmatrix} \in \Phi : |d_1| = |d_2|, d_1\overline{d}_2 + \overline{d}_1 d_2 = 0 \right\}$

$\Phi_s = \left\{ \Theta \in \Phi : P(X_t^\Theta = x) = P(X_t^\Theta = -x) \text{ for any } t \in Z_+ \text{ and } x \in Z \right\}$

$$\Phi_0 = \{\Theta \in \Phi : E(X_t^\Theta) = 0, \text{ for any } t \in Z_+\}$$

where $Z_+$ represents the set of positive integers and $Z$ represents the integers. It is noted that if $d_1 d_2 \neq 0$, then $\bar{d}_1 d_2 + d_1 \bar{d}_2 = 0$ imply that $d_1$ and $d_2$ are orthogonal. If $\Theta \in \Phi_s$, then the distribution of $X_t^\Theta$ is symmetric for any $t \in Z_+$ and the authors concluded the following theorem

**Theorem 5.1:** For $H*\left(\dfrac{1}{2}, \dfrac{1}{2}\right)$ we have $\Phi_\perp = \Phi_s = \Phi_0$.

In order for the theorem to hold the inclusion $\Phi_\perp \subseteq \Phi_s \subseteq \Phi_0 \subseteq \Phi_\perp$ needs to be proven. However, their proof of the inclusion $\Phi_0 \subseteq \Phi_\perp$ lacks rigor. To show this inclusion based on their class of initial quibit states, we need to show that if $\Theta \in \Phi_0$, then $\Theta \in \Phi_\perp$. However if $\Theta \in \Phi_0$, then, it is easily seen that $E(X_t^\Theta) = 0$, but this does not necessarily imply that $\Theta \in \Phi_\perp$. In particular the authors have shown for $t \in \{1, \ldots 10\}$ we have $E(X_t^\Theta) = -a_t \left(|d_1|^2 - |d_2|^2\right) - b_t \left(d_1 \bar{d}_2 + \bar{d}_1 d_2\right)$ where $a_1 = 0$, $a_2 = 0$, $a_3 = \dfrac{1}{2}$, $a_4 = 1$, $a_5 = \dfrac{9}{8}$, $a_6 = \dfrac{5}{4}$, $a_7 = \dfrac{27}{16}$, $a_8 = \dfrac{17}{8}$, $a_9 = \dfrac{293}{128}$, $a_{10} = \dfrac{157}{64}$,

$b_1 = 1$, $b_2 = 1$, $b_3 = 1$, $b_4 = \dfrac{3}{2}$, $b_5 = 2$, $b_6 = \dfrac{17}{8}$, $b_7 = \dfrac{9}{4}$, $b_8 = \dfrac{43}{16}$, $b_9 = \dfrac{25}{8}$, $b_{10} = \dfrac{421}{128}$ and conjectured that $b_{t+1} = a_t + 1$ for any $t \geq 1$. It follows at once, irrespective of the coefficients $a_t$ and $b_t$ that we can establish the following : If $\Theta \in \Phi_0$, then $E(X_t^\Theta) = 0$ _if and only if_ $\Theta \in \Phi_\perp$. This then begs the following question: If the Konno-Namiki-Soshi conjecture is true and $\Theta \notin \Phi_\perp$, does there exist $a_t$ and $b_t$ both not zero such that $E(X_t^\Theta) = 0$? We prove the answer is _true_ and use it to provide the correct the class of initial quibit states for which the Theorem 5.1 holds and generalizes immediately if we replace $H^*\left(\frac{1}{2},\frac{1}{2}\right)$ with $H^*(p,q)$.

To see the conjecture, one can use the principle of mathematical induction on $t$. First their conjecture can be reformulated as follows: Is $b_2 - a_1 = b_3 - a_2 = \ldots = b_t - a_{t-1} = b_{t+1} - a_t = 1 \;\; \forall t \geq 1$?

If $t \in \{1,2\ldots,10\}$, the authors have shown the answer is affirmative. It follows that the basis step is automatic. Now we assume for $t = k$ that $b_2 - a_1 = b_3 - a_2 = \ldots = b_k - a_{k-1} = b_{k+1} - a_k = 1$, then the inductive hypothesis imply we also have the following

$b_2 - a_1 = b_3 - a_2 = \ldots = b_k - a_{k-1} = b_{k+1} - a_k = b_{k+2} - a_{k+2} = 1$, so the equality holds for $t = k+1$, and the conjecture is resolved. Now we show that if $\Theta \notin \Phi_\perp$, then there exist $a_t$ and $b_t$ both not zero such that $E(X_t^\Theta) = 0$. Notice since the conjecture is true it follows that we can write

$E(X_t^\Theta) = L - b_{t+1}L - b_t Q$, where $L = |d_1|^2 - |d_2|^2$ and $Q = d_1 \bar{d}_2 + \bar{d}_1 d_2$. Now if $E(X_t^\Theta) = 0$, then this implies we have $b_{t+1}L + b_t Q = L$. Put $b_t^{homogeneous} = X^t$, and consider the related homogeneous equation $b_{t+1}L + b_t Q = 0$, it is easy to check that its solution, $b_t^{homogeneous}$, is given by

$$b_t^{homogeneous} = (-1)^t \left(\frac{Q}{L}\right)^t.$$

Now for $b_{t+1}L + b_t Q = L$ itself we put $b_t^{constant} = X$, and it easy to see that the associated solution is

$b_t^{constant} = \frac{L}{L+Q}$, thus $b_t = b_t^{homogeneous} + b_t^{constant} = (-1)^t \left(\frac{Q}{L}\right)^t + \frac{L}{L+Q}$ which is is the solution to

$b_{t+1}L + b_t Q - L = 0$. Since the left hand side of $b_{t+1}L + b_t Q - L = 0$ is $E(X_t^\Theta)$, we have found $b_t$ and $a_t = b_{t+1} - 1$ both not zero such that $E(X_t^\Theta) = 0$.

Now we give the correct class of initial quibit states for which Theorem 5.1 is valid and generalizes if we replace $H^*\left(\frac{1}{2},\frac{1}{2}\right)$ with $H^*(p,q)$. The class of initial quibit states are as follows:

$$\Phi_s = \{\Theta \in \Phi : P(X_t^\Theta = x) = P(X_t^\Theta = -x) \text{ for any } t \in Z_+ \text{ and } x \in Z\}$$

$$\Phi_\perp = \left\{\Theta = \begin{bmatrix} d_1 \\ d_2 \end{bmatrix} \in \Phi : |d_1| = |d_2|, d_1 \bar{d}_2 + \bar{d}_1 d_2 = 0\right\}$$

$$\Phi_0 = \Phi_0' \cup \Phi_0'', \text{ where } \Phi_0 = \{\Theta \in \Phi : E(X_t^\Theta) = 0\}, \Phi_0' = \{\Theta \in \Phi_\perp : E(X_t^\Theta) = 0\},$$

$$\Phi_0'' = \{\Theta \notin \Phi_\perp : E(X_t^\Theta) = 0\}.$$

In order to prove Theorem 5.1 the following Lemma is needed which is essentially Lemma 1 in [18] upon making the necessary change in notation so we omit the proof here.

**Lemma 5.2:** Let $J = \begin{bmatrix} 0 & -1 \\ 1 & 0 \end{bmatrix}$. For any $x \in Z$, $t \in Z_+$, we have

$$\zeta_x^t(\Theta) = \begin{cases} (-1)^t i J \zeta_{-x}^t(\Theta), & \text{if } \Theta = \sqrt{\frac{1}{2}} e^{i\theta} \begin{bmatrix} 1 \\ i \end{bmatrix} \\ (-1)^t (-i) J \zeta_{-x}^t(\Theta), & \text{if } \Theta = \sqrt{\frac{1}{2}} e^{i\theta} \begin{bmatrix} 1 \\ -i \end{bmatrix} \end{cases}, \text{ where } \theta \in [0, 2\pi)$$

**Proof of Theorem 5.1:** We need to show $\Phi_\perp \subseteq \Phi_s \subseteq \Phi_0$ and $\Phi_0 \subseteq \Phi_s \subseteq \Phi_\perp$ for equality to hold. Equivalently we can show the following inclusion $\Phi_\perp \subseteq \Phi_s \subseteq \Phi_0 \subseteq \Phi_\perp$.

We first show the inclusion $\Phi_s \subseteq \Phi_0$. Note that if $\Theta \in \Phi_s$, then $P(X_t^\Theta = x) = P(X_t^\Theta = -x)$ and so we can deduce that $\sum_x x P(X_t^\Theta = x) - \sum_{-x}(-x) P(X_t^\Theta = -x) = 0$, that is, $E(X_t^\Theta) = 0$.

This implies that $\Theta \in \Phi_0$ and the inclusion $\Phi_s \subseteq \Phi_0$ is complete.

We now show the inclusion $\Phi_\perp \subseteq \Phi_s$. First we note that for $\theta_1 \neq \theta_2$ where

$\theta_1, \theta_2 \in [0, 2\pi)$, it is easy to check that $\sqrt{\frac{1}{2}} e^{i\theta_1} \begin{bmatrix} 1 \\ i \end{bmatrix}$ and $\sqrt{\frac{1}{2}} e^{i\theta_2} \begin{bmatrix} 1 \\ -i \end{bmatrix}$ are in $\Phi_\perp$. Since the matrix

$A = \begin{bmatrix} V_1 & V_2 \end{bmatrix}$ have linearly independent columns where $V_1 = \sqrt{\frac{1}{2}} e^{i\theta_1} \begin{bmatrix} 1 \\ i \end{bmatrix}$ and $V_2 = \sqrt{\frac{1}{2}} e^{i\theta_2} \begin{bmatrix} 1 \\ -i \end{bmatrix}$, it follows that $\Phi_\perp = span\left\{ \sqrt{\frac{1}{2}} e^{i\theta_1} \begin{bmatrix} 1 \\ i \end{bmatrix}, \sqrt{\frac{1}{2}} e^{i\theta_2} \begin{bmatrix} 1 \\ -i \end{bmatrix} \right\}$. So if $\Theta \in \Phi_\perp$, then from Lemma 2.14 we can

deduce the following

$$P(X_t^\Theta = x) = |\zeta_x^t(\Theta)|^2 = \zeta_x^t(\Theta) \zeta_x^t(\Theta)^* = ((-1)^n i J \zeta_{-x}^t(\Theta))((-1)^n i^* J^* \zeta_{-x}^t(\Theta)^*) = \zeta_{-x}^t(\Theta) \zeta_{-x}^t(\Theta)^*$$
$$= |\zeta_{-x}^t(\Theta)|^2 = P(X_t^\Theta = -x)$$

which implies that $\Theta \in \Phi_s$ and the inclusion $\Phi_\perp \subseteq \Phi_s$ is complete. Now we show the inclusion

$\Phi_0 \subset \Phi_\perp$. If $\Theta \in \Phi_0$, then $E(X_t^\Theta) = 0$, but $\Phi_0 = \Phi_0' \cup \Phi_0''$, where $\Phi_0' = \{\Theta \in \Phi_\perp : E(X_t^\Theta) = 0\}$

and $\Phi_0'' = \{\Theta \notin \Phi_\perp : E(X_t^\Theta) = 0\}$ so $\Theta \in \Phi_0' \cup \Phi_0''$ and the definition of $\Phi_0$ implies that $\Theta \in \Phi_\perp$, and the inclusion $\Phi_0 \subset \Phi_\perp$ is complete.

Thus we immediately see that Theorem 5.1 generalizes if we replace $H^*\left(\frac{1}{2}, \frac{1}{2}\right)$ with $H^*(p, q)$. To see the generalization we first generalize Lemma 5.2 as follows.

**Lemma 5.3:** Let $J = \begin{bmatrix} 0 & -1 \\ 1 & 0 \end{bmatrix}$. For any $x \in Z$, $t \in Z_+$, we have

$$\zeta_x^t(\Theta) = \begin{cases} (-1)^t i J \zeta_{-x}^t(\Theta), & \text{if } \Theta = \sqrt{p} e^{i\theta} \begin{bmatrix} 1 \\ i \end{bmatrix} \\ (-1)^t (-i) J \zeta_{-x}^t(\Theta), & \text{if } \Theta = \sqrt{p} e^{i\theta} \begin{bmatrix} 1 \\ -i \end{bmatrix} \end{cases}, \text{ where } \theta \in [0, 2\pi)$$

**Proof:** The proof is essentially the same as Lemma 5.2 or Lemma 1 in [18], upon replacing the matrix $P$ with $P = \sqrt{p}\begin{bmatrix} 1 & 0 \\ 0 & 0 \end{bmatrix} + \sqrt{q}\begin{bmatrix} 0 & 1 \\ 0 & 0 \end{bmatrix}$ and the matrix $Q$ with

$Q = \sqrt{q}\begin{bmatrix} 0 & 0 \\ 1 & 0 \end{bmatrix} + \sqrt{p}\begin{bmatrix} 0 & 0 \\ 0 & -1 \end{bmatrix}$, and making the necessary change in notation.

Thus to see the generalized form of Theorem 5.1, we essentially use Lemma 5.3 in the proof of Theorem 5.1, and the generalized form of Theorem 5.1 follows which gives the necessary and sufficient condition for symmetry in the one dimensional model considered in this paper.

### 5.2. Necessary and Sufficient Condition in the Two Dimensional Model

Recall for the two dimensional model considered in this paper, that the time evolution is given by the transformation

$$H^{**}(p,q) = \begin{bmatrix} p & \sqrt{pq} & \sqrt{pq} & q \\ \sqrt{pq} & -p & q & -\sqrt{pq} \\ \sqrt{pq} & q & -p & -\sqrt{pq} \\ q & -\sqrt{pq} & -\sqrt{pq} & p \end{bmatrix}; \; q = 1-p$$

Recall that the direction of motion of the particle depends on its chirality state. In two dimensions we have four chirality states: left, right, up, or down. The evolution of the two-dimensional Hadamard walk proceeds as follows: At each time step, if the particle has left chirality, it moves one step to the left, and if it has right chirality, it moves one step to the right, and if the chirality state is up or down then it moves up or down, respectively. In particular $H^{**}(p,q)$ acts on four chirality states $|L\rangle$, $|R\rangle$, $|U\rangle$, and $|D\rangle$, where L, R, U and D refer to the left, right, up, and down chirality states respectively, as follows:

$|L\rangle \mapsto p|L\rangle + \sqrt{pq}(|R\rangle + |D\rangle) + q|U\rangle$

$|R\rangle \mapsto \sqrt{pq}(|L\rangle - |U\rangle) - p|R\rangle + q|D\rangle$

$|D\rangle \mapsto \sqrt{pq}(|L\rangle - |U\rangle) + q|R\rangle - p|D\rangle$

$$|U\rangle \mapsto q|L\rangle - \sqrt{pq}(|R\rangle + |D\rangle) + p|U\rangle$$

Define

$$|L\rangle = \begin{bmatrix} 1 \\ 0 \\ 0 \\ 0 \end{bmatrix}, |R\rangle = \begin{bmatrix} 0 \\ 1 \\ 0 \\ 0 \end{bmatrix}, |D\rangle = \begin{bmatrix} 0 \\ 0 \\ 1 \\ 0 \end{bmatrix}, \text{and } |U\rangle = \begin{bmatrix} 0 \\ 0 \\ 0 \\ 1 \end{bmatrix}, \text{then we have}$$

$$H^{**}(p,q)|L\rangle \mapsto p|L\rangle + \sqrt{pq}(|R\rangle + |D\rangle) + q|U\rangle$$

$$H^{**}(p,q)|R\rangle \mapsto \sqrt{pq}(|L\rangle - |U\rangle) - p|R\rangle + q|D\rangle$$

$$H^{**}(p,q)|D\rangle \mapsto \sqrt{pq}(|L\rangle - |U\rangle) + q|R\rangle - p|D\rangle$$

$$H^{**}(p,q)|U\rangle \mapsto q|L\rangle - \sqrt{pq}(|R\rangle + |D\rangle) + p|U\rangle$$

Now we introduce the following matrices

$$P_L = \begin{bmatrix} p & \sqrt{pq} & \sqrt{pq} & q \\ 0 & 0 & 0 & 0 \\ 0 & 0 & 0 & 0 \\ 0 & 0 & 0 & 0 \end{bmatrix}, \quad P_R = \begin{bmatrix} 0 & 0 & 0 & 0 \\ \sqrt{pq} & -p & q & -\sqrt{pq} \\ 0 & 0 & 0 & 0 \\ 0 & 0 & 0 & 0 \end{bmatrix},$$

$$P_D = \begin{bmatrix} 0 & 0 & 0 & 0 \\ 0 & 0 & 0 & 0 \\ \sqrt{pq} & q & -p & -\sqrt{pq} \\ 0 & 0 & 0 & 0 \end{bmatrix}, \quad P_U = \begin{bmatrix} 0 & 0 & 0 & 0 \\ 0 & 0 & 0 & 0 \\ 0 & 0 & 0 & 0 \\ q & -\sqrt{pq} & -\sqrt{pq} & p \end{bmatrix}$$

with $H^{**}(p,q) = P_L + P_R + P_D + P_U$. Here $P_L$, $P_R$, $P_D$, $P_U$ represents the probability the particle moves left, right, down, and up respectively. By using $P_L$, $P_R$, $P_D$, and $P_U$ we define the dynamics of the generalized Hadamard walk in two dimensions by the following $(2N+1)^2 \times (2N+1)^2$ matrix

$\overline{H}_N^{**} : (C^4)^{4N^2+4N+1} \mapsto (C^4)^{4N^2+4N+1}$ by $\overline{H}_N^{**} = \overline{H}_N^* \otimes \overline{H}_N^*$, where the

$(2N+1) \times (2N+1)$ matrix $\overline{H}_N^* : (C^2)^{2N+1} \mapsto (C^2)^{2N+1}$ is given by

$$\overline{H}_N^* = \begin{bmatrix} 0_{2\times 2} & P & 0_{2\times 2} & \cdots & & \cdots & 0_{2\times 2} & Q \\ Q & 0_{2\times 2} & P & 0_{2\times 2} & \cdots & & & 0_{2\times 2} \\ 0_{2\times 2} & Q & 0_{2\times 2} & P & 0_{2\times 2} & \cdots & & 0_{2\times 2} \\ \vdots & \ddots & \ddots & \ddots & \ddots & \ddots & & \vdots \\ 0_{2\times 2} & \cdots & 0_{2\times 2} & Q & 0_{2\times 2} & P & & 0_{2\times 2} \\ 0_{2\times 2} & \cdots & & & 0_{2\times 2} & Q & 0_{2\times 2} & P \\ P & 0_{2\times 2} & \cdots & & & 0_{2\times 2} & Q & 0_{2\times 2} \end{bmatrix}, \text{ where } 0_{2\times 2} = \begin{bmatrix} 0 & 0 \\ 0 & 0 \end{bmatrix}, P = \begin{bmatrix} \sqrt{p} & \sqrt{q} \\ 0 & 0 \end{bmatrix}, \text{ and}$$

$Q = \begin{bmatrix} 0 & 0 \\ \sqrt{q} & -\sqrt{p} \end{bmatrix}$, and $C$ is the set of complex numbers. Note that the $(2N+1) \times (2N+1)$ matrix

$\overline{H}_N^* : (C^2)^{2N+1} \mapsto (C^2)^{2N+1}$ defined above is associated with the dynamics of the generalized

Hadamard walk in one dimension. In particular recall that the relationship between the transformations

in the one and two dimensional models is given by $H^{**}(p,q) = H^*(p,q) \otimes H^{**}(p,q)$. Now let

$$\Omega_{x,y}^t(\theta) = \begin{bmatrix} \zeta_{1,x,y}^t(\theta) \\ \zeta_{2,x,y}^t(\theta) \\ \zeta_{3,x,y}^t(\theta) \\ \zeta_{4,x,y}^t(\theta) \end{bmatrix} = \zeta_{1,x,y}^t(\theta)|L\rangle + \zeta_{2,x,y}^t(\theta)|R\rangle + \zeta_{3,x,y}^t(\theta)|D\rangle + \zeta_{4,x,y}^t(\theta)|U\rangle \in C^4$$

be the four component vector of amplitudes of the particle being at site $(x, y)$ at time $t$, where the first, second, third and fourth components of the column vector correspond to the chirality states left, right, down and up, respectively, and let $\Omega^t(\theta) =^T [\Omega_{-N}^t(\theta), \Omega_{-(N-1)}^t(\theta), \ldots, \Omega_{-N}^t(\theta)] \in (C^4)^{4N^2+4N+1}$ be

the quibit states at time $t$, where $T$ means the transposed operator. The initial quibit state is given by

$$\Omega^0(\theta) =^T \left[ \overbrace{0_{4\times 1},\ldots 0_{4\times 1}}^{2N^2+2N},\theta,\overbrace{0_{4\times 1},\ldots 0_{4\times 1}}^{2N^2+2N} \right] \in (C^4)^{4N^2+4N+1}, \text{ where } 0_{4\times 1} = \begin{bmatrix} 0 \\ 0 \\ 0 \\ 0 \end{bmatrix} \text{ and } \theta = \begin{bmatrix} k_1 \\ k_2 \\ k_3 \\ k_4 \end{bmatrix} \text{ with }$$

$\sum_{i=1}^{4}|k_i|^2 = 1$. The following equation defines the time evolution of the generalized two-dimensional Hadamard walk

$$\left(\overline{H}_N^{**}\Omega^t(\theta)\right)_{x,y} = P_R\,\Omega_{x-1,y}^t(\theta) + P_L\,\Omega_{x+1,y}^t(\theta) + P_U\,\Omega_{x,y-1}^t(\theta) + P_D\,\Omega_{x,y+1}^t(\theta), \text{ where }$$

$\left(\Omega^t(\theta)\right)_{x,y} = \Omega_{x,y}^t(\theta)$. Note that $P_R, P_L, P_U, P_D$ satisfy

$$P_R P_R^* + P_L P_L^* + P_U P_U^* + P_D P_D^* + P_R^* P_R + P_L^* P_L + P_U^* P_U + P_D^* P_D = I_{4\times 4}, \text{ and also satisfy}$$

$$P_R P_L^* = P_L P_R^* = P_L^* P_R = P_R^* P_L = \begin{bmatrix} 0 & 0 & 0 & 0 \\ 0 & 0 & 0 & 0 \\ 0 & 0 & 0 & 0 \\ 0 & 0 & 0 & 0 \end{bmatrix} = P_U P_D^* = P_D P_U^* = P_D^* P_U = P_U^* P_D, \text{ where * means}$$

the adjoint operator. The above relations imply that $\overline{H}_N^{**}$ is also a unitary matrix. Define the set of initial quibit states as follows: $\Phi = \left\{ \theta = \begin{bmatrix} k_1 \\ k_2 \\ k_3 \\ k_4 \end{bmatrix} \in C^4 : \sum_{i=1}^{4}|k_i|^2 = 1 \right\}$. Now we define the probability distribution

of the generalized Hadamard walk in two dimensions starting from the initial quibit state $\theta \in \Phi$

by $P\left((X_t^\theta, Y_t^\theta) = (x,y)\right) = \left|\Omega_{x,y}^t(\theta)\right|^2$. Now we introduce the following class of initial quibit states akin to the one dimensional case.

$$\tilde{\Phi}_s = \left\{ \theta \in \Phi : \left|\Omega_{-x,y}^t(\theta)\right|^2 = \left|\Omega_{x,-y}^t(\theta)\right|^2 = \left|\Omega_{-x,-y}^t(\theta)\right|^2 = \left|\Omega_{x,y}^t(\theta)\right|^2 \text{ for any } t \in Z_+ \text{ and } (x,y) \in Z^2 \right\}$$

$$\Phi_{\perp}^{\approx} = \left\{ \theta = \begin{bmatrix} k_1 \\ k_2 \\ k_3 \\ k_4 \end{bmatrix} \in \Phi : |k_1|^2 = |k_2|^2 = |k_3|^2 = |k_4|^2, \sum_{\substack{i,j \\ i \neq j}} k_i \bar{k}_j = 0 \right\}$$

$$\Phi_0^{\approx} = \left\{ \theta \in \Phi_{\perp}^{\approx} : E(X_t^\theta Y_t^\theta) = 0 \text{ for any } t \in Z_+ \right\} \cup \left\{ \theta \notin \Phi_{\perp}^{\approx} : E(X_t^\theta Y_t^\theta) = 0 \text{ for any } t \in Z_+ \right\}.$$

Akin to the one dimensional case it also follows that if $k_1 k_2 k_3 k_4 \neq 0$, then $\sum_{\substack{i,j \\ i \neq j}} k_i \bar{k}_j = 0$, implies that, $k_1, k_2, k_3, k_4$ are orthogonal. For $\theta \in \Phi_s^{\approx}$, the distribution of the generalized Hadamard walk in two dimensions is also symmetric, and the following theorem below establishes the necessary and sufficient condition.

**Theorem 5.4:** $\Phi_s^{\approx} = \Phi_{\perp}^{\approx} = \Phi_0^{\approx}$

The proof of the above theorem is similar in nature to the proof of generalized form of Theorem 5.1 which was obtained by replacing $H^*\left(\frac{1}{2}, \frac{1}{2}\right)$ with $H^*(p, q)$, and involves invoking the following Lemma whose proof is also similar to that of the generalized form of Lemma 5.3.

**Lemma 5.5:** Let $J^{\approx} = \begin{bmatrix} 0 & 0 & 0 & 1 \\ 0 & 0 & -1 & 0 \\ 0 & -1 & 0 & 0 \\ 1 & 0 & 0 & 0 \end{bmatrix}$. For any $(x, y) \in Z^2$, $t \in Z_+$, and $\gamma \in [0, 2\pi)$ we have

$$\Omega_{x,y}^{t}(\theta) = \begin{cases} (-1)^{n} i J^{*} \Omega_{-x,-y}^{t}(\theta); \text{ if } \theta = pe^{2i\gamma} \begin{bmatrix} 1 \\ i \\ i \\ -1 \end{bmatrix} \\ \\ (-1)^{n} (-i) J^{*} \Omega_{-x,-y}^{t}(\theta); \text{ if } \theta = pe^{2i\gamma} \begin{bmatrix} 1 \\ -i \\ -i \\ -1 \end{bmatrix} \end{cases}, \quad \Omega_{x,-y}^{t}(\theta) = \begin{cases} (-1)^{n} i J^{*} \Omega_{-x,-y}^{t}(\theta); \text{ if } \theta = pe^{2i\gamma} \begin{bmatrix} 1 \\ i \\ i \\ -1 \end{bmatrix} \\ \\ (-1)^{n} (-i) J^{*} \Omega_{-x,-y}^{t}(\theta); \text{ if } \theta = pe^{2i\gamma} \begin{bmatrix} 1 \\ -i \\ -i \\ -1 \end{bmatrix} \end{cases}$$

## 6. Concluding Remarks

We have obtained weak convergence theorems in the long-time limit of the walker's pseudo-velocity in both one and two dimensions for the Hadamard-walk model considered in this paper. Theoretical criterion is given for *localization* and *symmetrization* in both the one and two dimensional models. It is an interesting problem to study symmetrization via a variant of the Quantum Pascal Triangle as considered by [18] in the case of the unbiased Hadamard walk in one-dimension.